\documentclass[a4paper,10pt,twoside]{cpc-hepnp}

\usepackage{multicol}
\usepackage{graphicx}
\usepackage{booktabs}
\usepackage{amssymb,bm,mathrsfs,bbm,amscd}
\usepackage[tbtags]{amsmath}
\usepackage{lastpage}
\begin{document}

\fancyhead[co]{\footnotesize Submitted to 'Chinese Physics C'}

\footnotetext[0]{Received \today}

\title{A large area plastic scintillation detector with 4-corner-readout\thanks{Supported by National Natural Science
Foundation of China (U1332207, 11405242) }}

\author{%
      Tang Shu-Wen$^{1}$
\quad Yu Yu-Hong$^{1;1)}\email{yuyuhong@impcas.ac.cn}$
\quad Zhou Yong$^{1;2}$
\quad Sun Zhi-Yu$^{1}$\\
\quad Zhang Xue-Heng$^{1}$
\quad Wang Shi-Tao$^{1}$
\quad Yue Ke$^{1}$
\quad Liu Long-Xiang$^{1}$\\
\quad Fang Fang$^{1}$
\quad Yan Duo$^{1;2}$
\quad Sun Yu$^{1;2}$
\quad Wang Zhao-Ming$^{1;2}$
}
\maketitle

\address{%
$^1$ Institute of Modern Physics, Chinese Academy of Sciences, Lanzhou 730000, China\\
$^2$ University of Chinese Academy of Sciences, Beijing 100049, China\\
}

\begin{abstract}
A 760 $\times$ 760 $\times$ 30 mm$^3$ plastic scintillation detector viewed by photomultiplier tubes (PMTs) from four corners has been developed, and the detector has been tested with cosmic rays and $\gamma$ rays. A position-independent effective time T$_{eff}$ has been found, indicating this detector can be used as a TOF detector. The hit position can also be reconstructed by the time from four corners. A TOF resolution of 236~ps and a position resolution of 48~mm have been achieved, and the detection efficiency has also been investigated.
\end{abstract}



\begin{multicols}{2}

\section{Introduction}
With the developments of the time-of-flight (TOF) technique, plastic scintillator is widely used in the modern nuclear physics laboratories, due to its fast time response, relatively low cost and easy fabricating into a variety of forms useful for many experiments. Recently, numerous large plastic scintillation spectrometers for detection of charged particles and neutrons with energy from tens to hundreds MeV per nucleon have been reported~\cite{Baldo04, Kouznetsova02, Grabmayr98, Blaich92, Yu09}. Such detectors are normally placed far away ($>$10~m) downstream the reaction target. As a result, it is essential to use a large size detector to cover enough acceptance. One of the typical design is to use a modular array. It consists of a number of long plastic scintillator bars which is viewed by two PMTs at both ends. This type of detectors always obtain good time resolution (typically $\sigma$=150~ps) and acceptable position resolution (typically $\sigma$=3~cm, along the long direction of the bar). Besides, the multi-hit events are easy to be handled because of the segmented construction. However, one big problem always exists in this modular design. A large number of bars, equivalently a large number of PMTs and associated electronics channels, means a very high cost and a lot of time-consuming calibration work. One natural way to solve this problem is to use a large volume of plastic scintillator viewed by minimum number of PMTs.

In this paper, a large area scintillation detector has been developed and the detailed performance tested with cosmic rays and $\gamma$ rays will be also presented.

\section{Detector Construction}
The schematic layout of the detector is shown in Fig.~\ref{fig1}. The sensitive area is a 760 $\times$ 760~mm$^2$ square \begin{center}
\includegraphics[width=6cm]{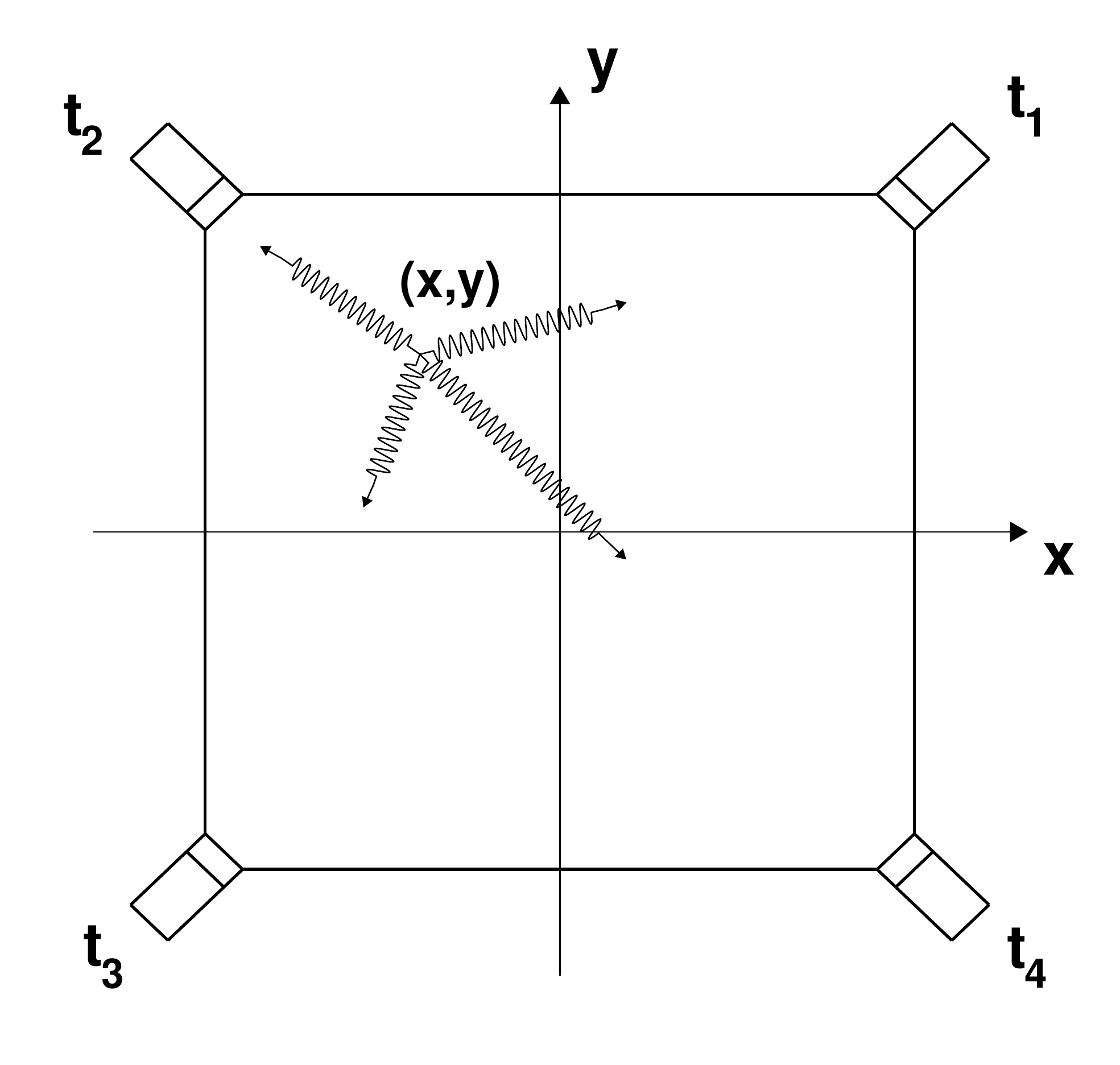}
\figcaption{\label{fig1} Schematic layout of the scintillation detector. (x,y) is the hit position of an incident particle, $t_1$, $t_2$, $t_3$, $t_4$ are the most probable propagation times reaching four PMTs. }
\end{center}
and the thickness is 30~mm. The plastic scintillator EJ-200, purchased from ELJEN Corporation, is chosen as the detection material due to its fast timing (2.1~ns decay time), long optical attenuation length (4~m in bulk) and high scintillation efficiency ($\approx$ 10$^4$ photon / MeV). The square detector was cut off from each corner to form a plane with a cross section of 42.5 $\times$ 30~mm$^2$. Each plane was glued to a light guide which is mechanically coupled to a Hamamatsu R7724 PMT by means of optical silicone grease. To improve the transmission efficiency of the scintillation light, each surface of the scintillator as well as the light guides was finely polished by the manufacturer. The scintillators and the light guides were first carefully wrapped in a layer of 0.15~mm tyvek paper, leaving no air gap to provide well reflection, and then wrapped with black foam and black tape for light proofing. A high voltage divider circuit was mounted directly behind each PMT. For mechanical protection and light shielding, the readout PMT and its voltage divider were mounted in a PMT housing.

\section{Performance of the detector}
\subsection{Time performance}
To investigate the time characteristics of the large area scintillation detector, there were 49 positions in $7\times7$ grid points chosen for study, and every two adjacent grid points had an interval of 10~cm. Cosmic ray muons are very suitable for us to study the performance of the square detector as muons loose a fairly constant energy when they pass through the scintillator material. The greatest challenge is the rate of muons in each measured grid point is too slow, and it will spend too much time to accumulate enough events. One way to deal with this problem is to test with high rate $\gamma$ rays for all the 49 points instead, and then do comparison test with cosmic rays at some certain points. Fig.~\ref{fig2} shows the schematic layout of the two tests. During the $\gamma$ rays test (Fig.~\ref{fig2}a), a lead collimator with the height of 10~cm and the hole size of 5 mm in diameter was used for position determination. One of the paired $\gamma$ rays from a $^{60}Co$ source went into the upper small trigger scintillator (S1), and the other passed through the collimator hole and then was detected by the large area scintillation detector. During the cosmic rays test (Fig.~\ref{fig2}b), three small trigger scintillators (S1, S2 and S3) were applied to electronically collimate the muons with triple coincidence. Each of the trigger detectors was constructed from a piece of 2 $\times$ 2 $\times$ 1 cm$^3$ BC-408 plastic scintillator coupled with a Hamamatsu 
\begin{center}
\includegraphics[width=6cm]{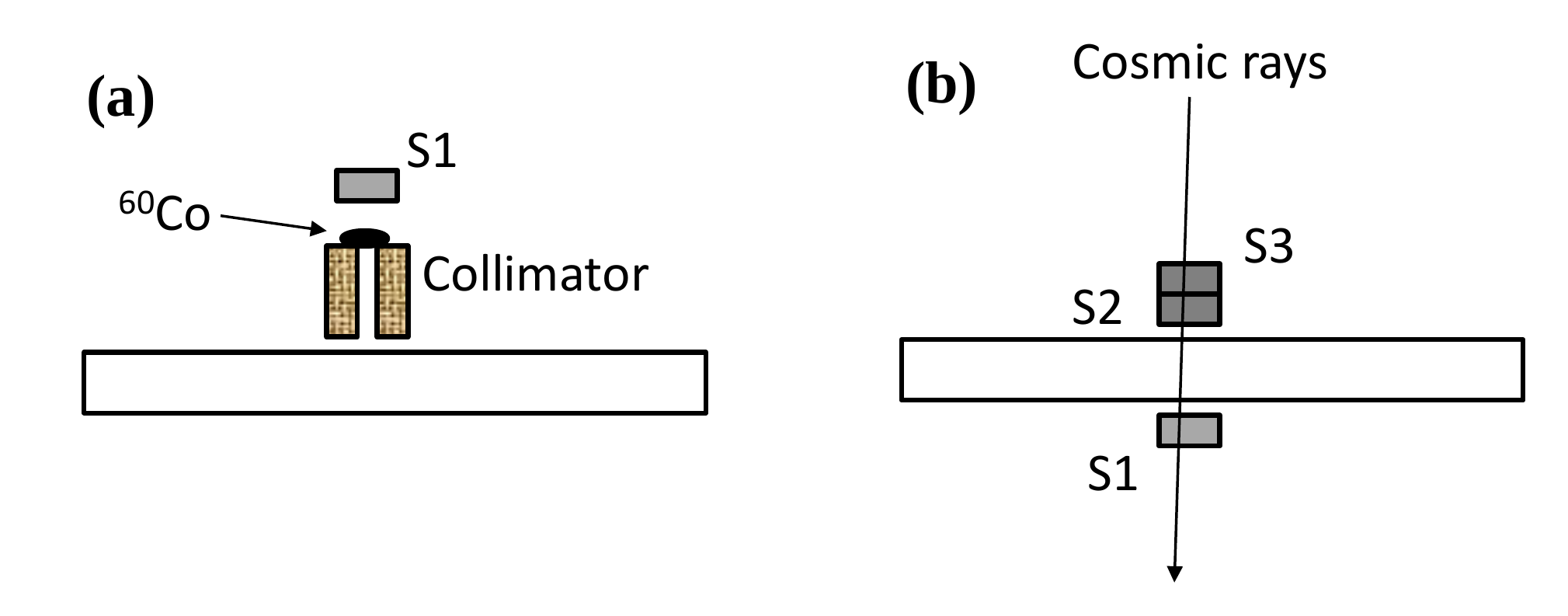}
\figcaption{\label{fig2} Schematic layout of the detector tests. (a) test with a $^{60}Co~ \gamma$ source. (b) test with cosmic rays.  }
\end{center}
R7525 PMT. Different positions of the large area scintillation detector could be tested by moving the trigger detectors. Both of the tests adopted CAMAC-CC32 system as data acquisition. The time information from all of the PMT signals was recorded in a Phillips 7186 TDC, and the start timing was made by the trigger detectors.

Rather strong position-dependent time results were observed for all PMTs of the large area scintillation detector. Fig.~\ref{fig3} shows the time information from one PMT signal varies with the distance to the tested position, which can be described with Eq.~\ref{eq1}:
\begin{equation}
\label{eq1}
    t=\delta t + \frac{l}{v_{eff}}
\end{equation}
where $l$ is the distance from the tested position to the PMT, $v_{eff}$ is the effective light velocity, and $\delta t$ is the time delayed by PMT and electronics. It is easy to obtain $v_{eff}$=15 cm / ns from Fig.~\ref{fig3}. \begin{center}
\includegraphics[width=6cm]{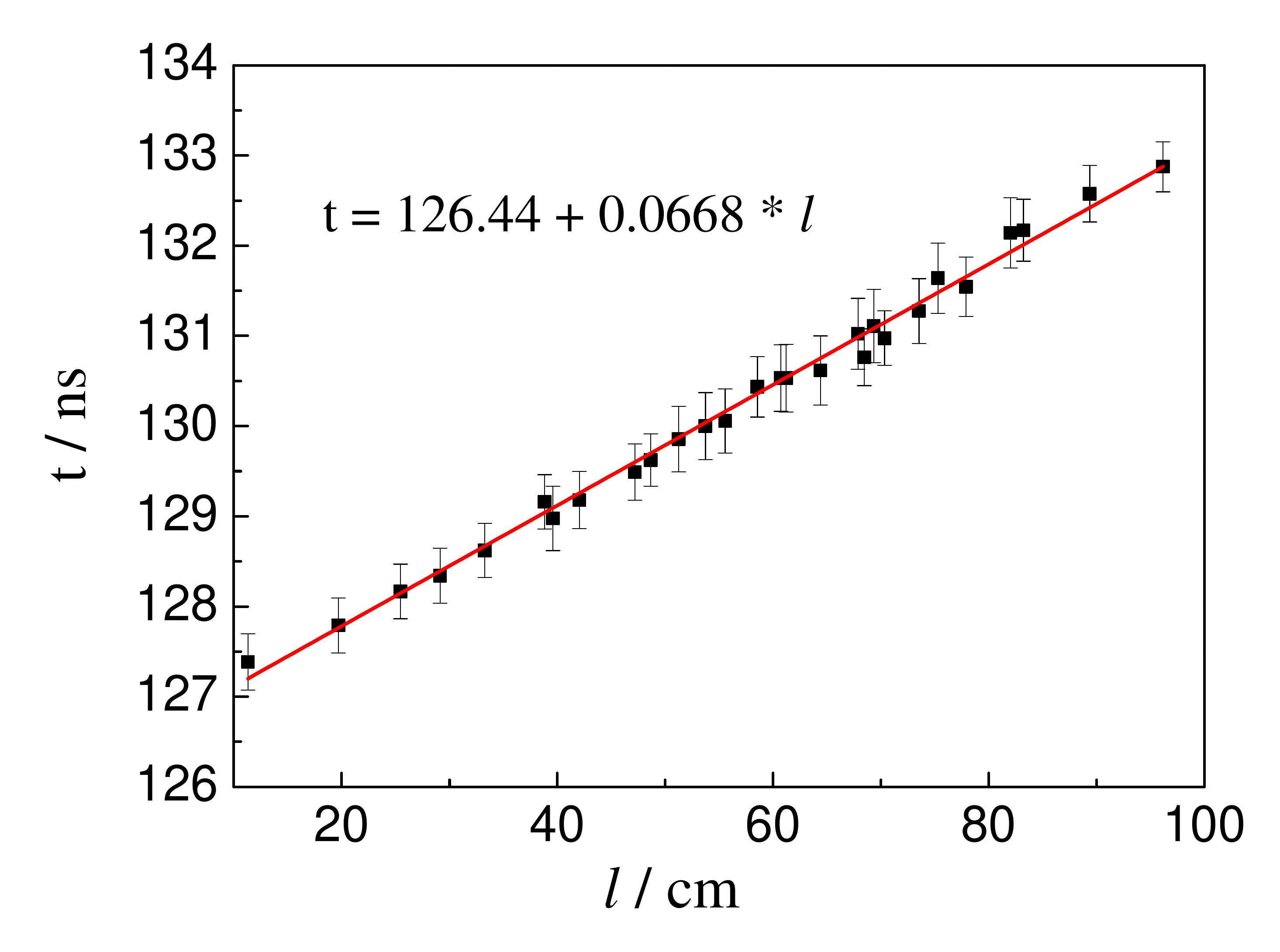}
\figcaption{\label{fig3} Position-dependent time for one PMT of the detector. }
\end{center}

With the value of $v_{eff}$, the effective time $T_{eff}$ can be obtained from Eq.~\ref{eq2} according to Ref.~\cite{Annand87}:
\begin{equation}
\label{eq2}
    T_{eff} = \frac{1}{4}\sum^4_{i=1} T_i - \frac{v_{eff}}{16a} [(T_3 - T_1)^{2} + (T_4 - T_2)^2]
\end{equation}
where a is the half diagonal length of the detector, $T_i$ are output times from four corner PMTs, and the subscripts 1, 3 and 2, 4 refer to diagonally opposite corner (see Fig.~\ref{fig1}). $T_i$ can be calculated from $T_i = t_i - \delta t_i$, where $t_i$ and $\delta t_i$ are respectively corresponding to TDC time and zero time obtained from PMT$_i$. The results of $T_{eff}$ for all tested positions are plotted in Fig.~\ref{fig4}. Each of the results is filled in a curly brace and drawn at the corresponding tested grid point. The upper value in the curly brace is $T_{eff}$ and the lower
value is the time resolution ($\sigma$), both of them are in nanosecond. The average $T_{eff}$ value is 3.471~ns with a RMS of 111~ps, and the average $\sigma$ value is 375~ps with a RMS of 55~ps. Considering the value of time resolution, we can claim that $T_{eff}$ is position-independent in the whole detector range.
\begin{center}
\includegraphics[width=8cm]{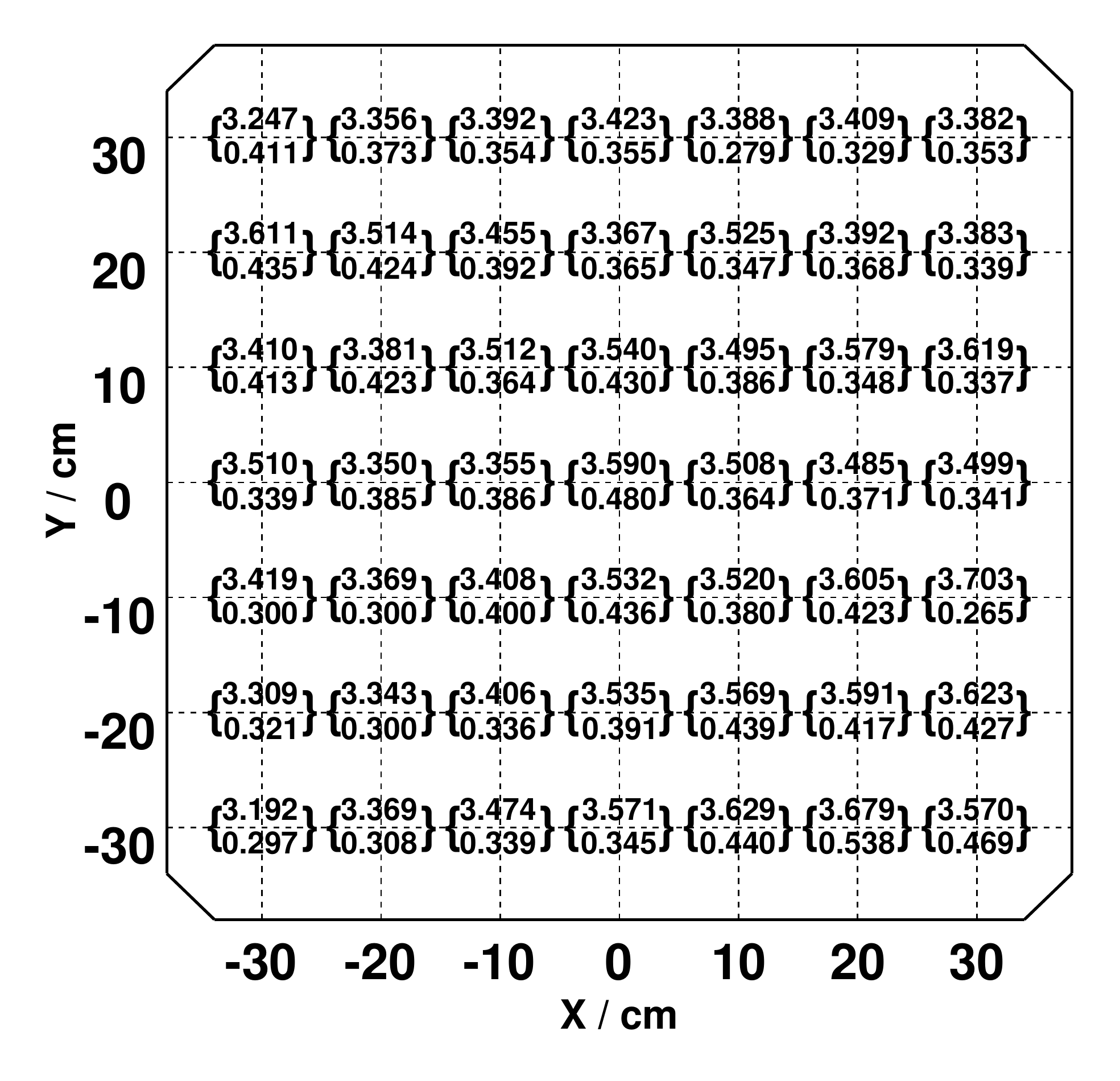}
\figcaption{\label{fig4} The effective time $T_{eff}$ and the time resolution ($\sigma$) at all grid positions tested with $\gamma$ rays, all values are in nanosecond. }
\end{center}

Cosmic-ray tests were made at three positions of the large detector, one in the center, another one in the bottom side and the last one in the left bottom corner. The comparison results show that the propagation time of the scintillation light from its generated position to each PMT of the detector is equivalent between cosmic-ray test and $\gamma$-ray test at every tested position, but the time resolutions from cosmic-ray tests are all better for their certain energy loss in the detector material. Fig.~\ref{fig5}a shows the $T_{eff}$ spectrum for cosmic-ray test and $\gamma$-ray test at the center position.
\begin{center}
\includegraphics[width=8cm]{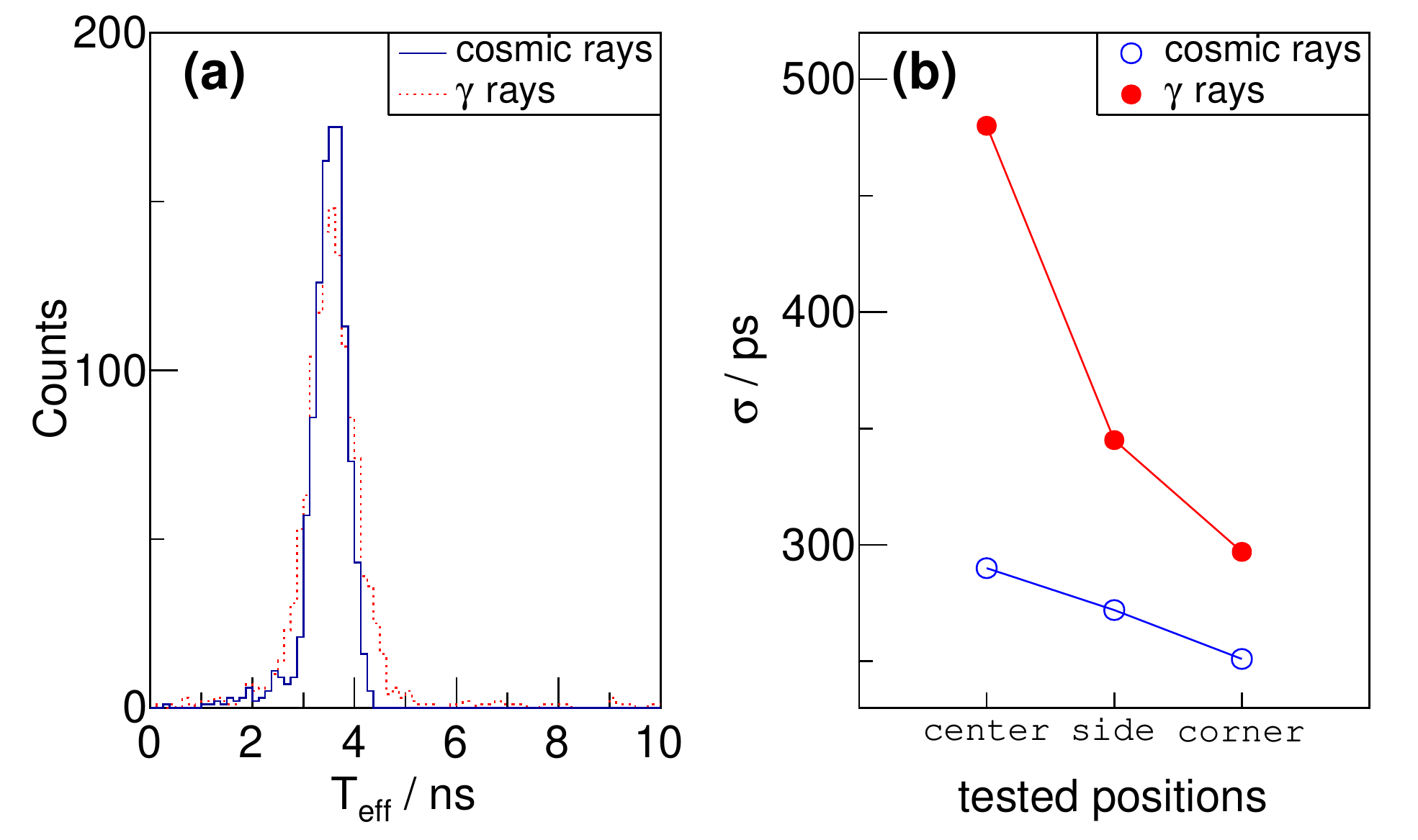}
\figcaption{\label{fig5} (a) T$_{eff}$ distribution for cosmic rays and $\gamma$ rays tests at the center position of the detector. (b) Time resolution for cosmic rays and $\gamma$ rays tests at three tested positions.
}
\end{center}
The time resolution from cosmic-ray tests at the three tested points are 290~ps, 272~ps and 251~ps respectively, and the comparison with corresponding $\gamma$ rays results is plotted in Fig~\ref{fig5}b. The average time resolution from cosmic rays is 271~ps which includes the resolution from the trigger detector. After eliminating the contribution of 133~ps from the trigger detector, the average time resolution is 236~ps. According to the time resolution results investigated with $\gamma$ rays, the average time resolution of the three positions can almost represent the average time resolution of the whole detector. Therefore, the time resolution for the large scintillation detector is about 236~ps for cosmic rays test.

\begin{center}
\includegraphics[width=6cm]{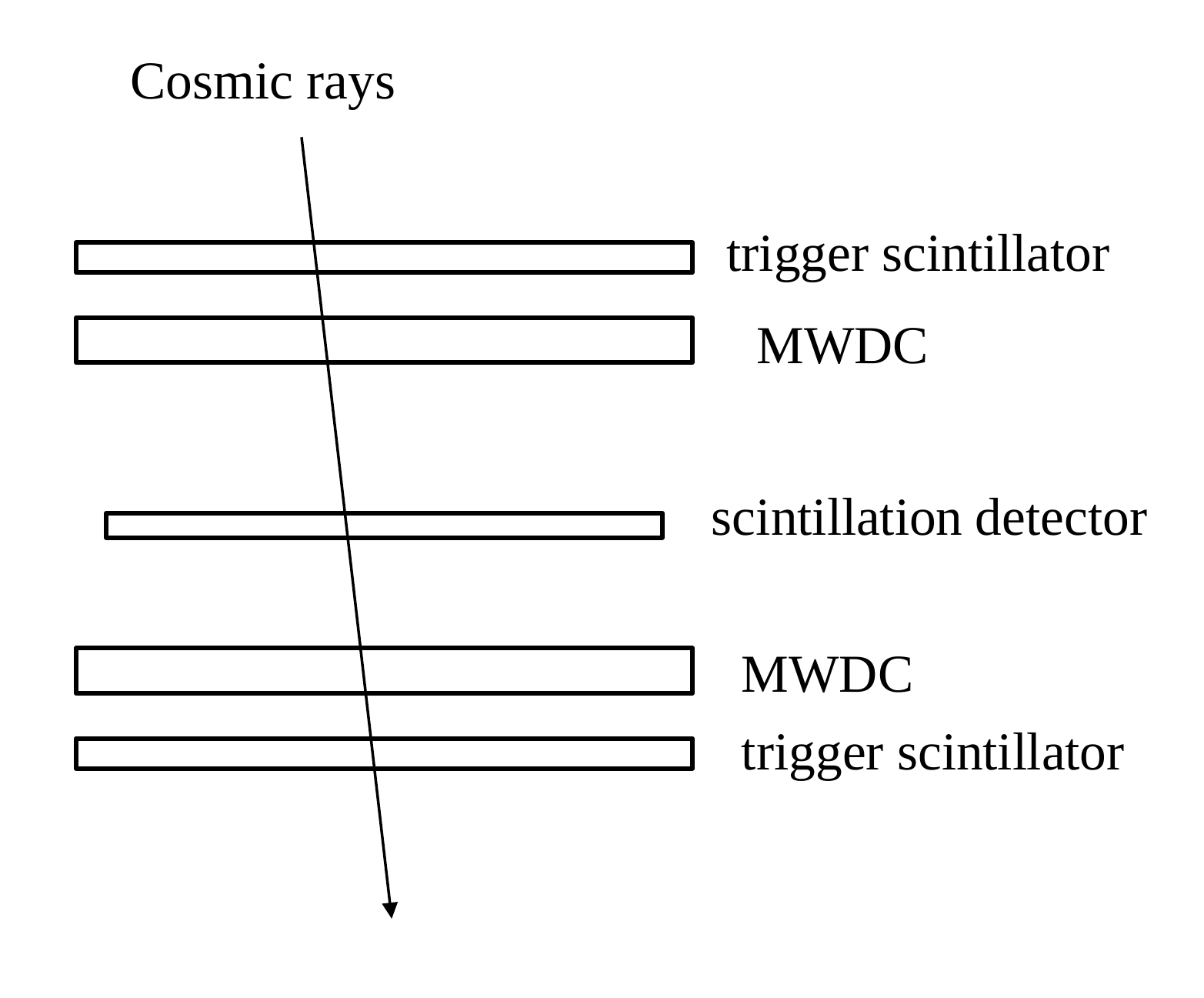}
\figcaption{\label{fig6} Schematic layout of position test with cosmic rays. }
\end{center}
\subsection{Position reconstruction}
The detector was tested with cosmic rays in a universal detector test platform, which consists of two multi-wire drift chambers (MWDCs) to determine the trajectory and two large area plastic scintillators as trigger detectors, the schematic layout is plotted in Fig.~\ref{fig6}. Each of the trigger detectors has an active area of 825 $\times$ 825 mm$^2$, and is also 4-corner-readout by PMTs. The active area of each MWDC is 830 $\times$ 830~mm$^2$ and the position resolution is below 1~mm. The detector to be tested was placed in the middle of the platform. If eight PMTs signals from both of the upper and lower trigger detectors were all measured at the meantime, a valid cosmic ray event was accepted. Then all PMT signals would be digitized in a 16-channel high precise HPTDC module~\cite{Zhao14}, and signals from the MWDC wires were recorded in several 128-channel HPTDC modules~\cite{Kang13}. As discussed in section 3.1, the arrival time of each single PMT time is dependent on the hit position. The time difference from the PMTs of the trigger detector can be deduced to be less than 8~ns according to $v_{eff}$ measured before. It makes no trouble for the signals from MWDCs as the drift time is very slow, but for the fast signal from scintillation detector, the waggling time of the trigger can not be negligible so this time is useless for start timing. However, the HPTDC modules can provide every timing moment by the precision of 24~ps with an internal clock. Therefore, it is unnecessary to use the start timing information. We can only use the four timing moments ($T_1$, $T_2$, $T_3$, $T_4$) from the corner PMTs to deduce the hit position $(x, y)$.

The time differences from diagonally opposite corners can be expressed with Eq.~\ref{eq3} and Eq.~\ref{eq4}:
\begin{eqnarray}
\label{eq3}
    T_{3}-T_{1}=\delta t_{31} + \frac{l_{3}-l_{1}}{v_{eff}}\\
    T_{4}-T_{2}=\delta t_{42} + \frac{l_{4}-l_{2}}{v_{eff}}
\label{eq4}
\end{eqnarray}
where $l_i$ (i=1, 2, 3, 4) is the distance from the hit position $(x,y)$ to corner $i$, $\delta t_{31}$ and $\delta t_{42}$ are constants determined by the electronics. If we substitute $l_i$ with corresponding expressions of $x$ and $y$ and solve the equations, the hit position will be obtained. Here we define it as "measured position", labeled as $x_m$ and $y_m$. The hit position can be also obtained from the two MWDCs through reconstruction of the trajectory of the cosmic ray muons. Because of the high position resolution of the MWDCs, the reconstructed hit position from the two auxiliary detector means the physical location of the incident rays. Here we define such position as "true position", which is labeled as $x_t$ and $y_t$.
\begin{center}
\includegraphics[width=6cm]{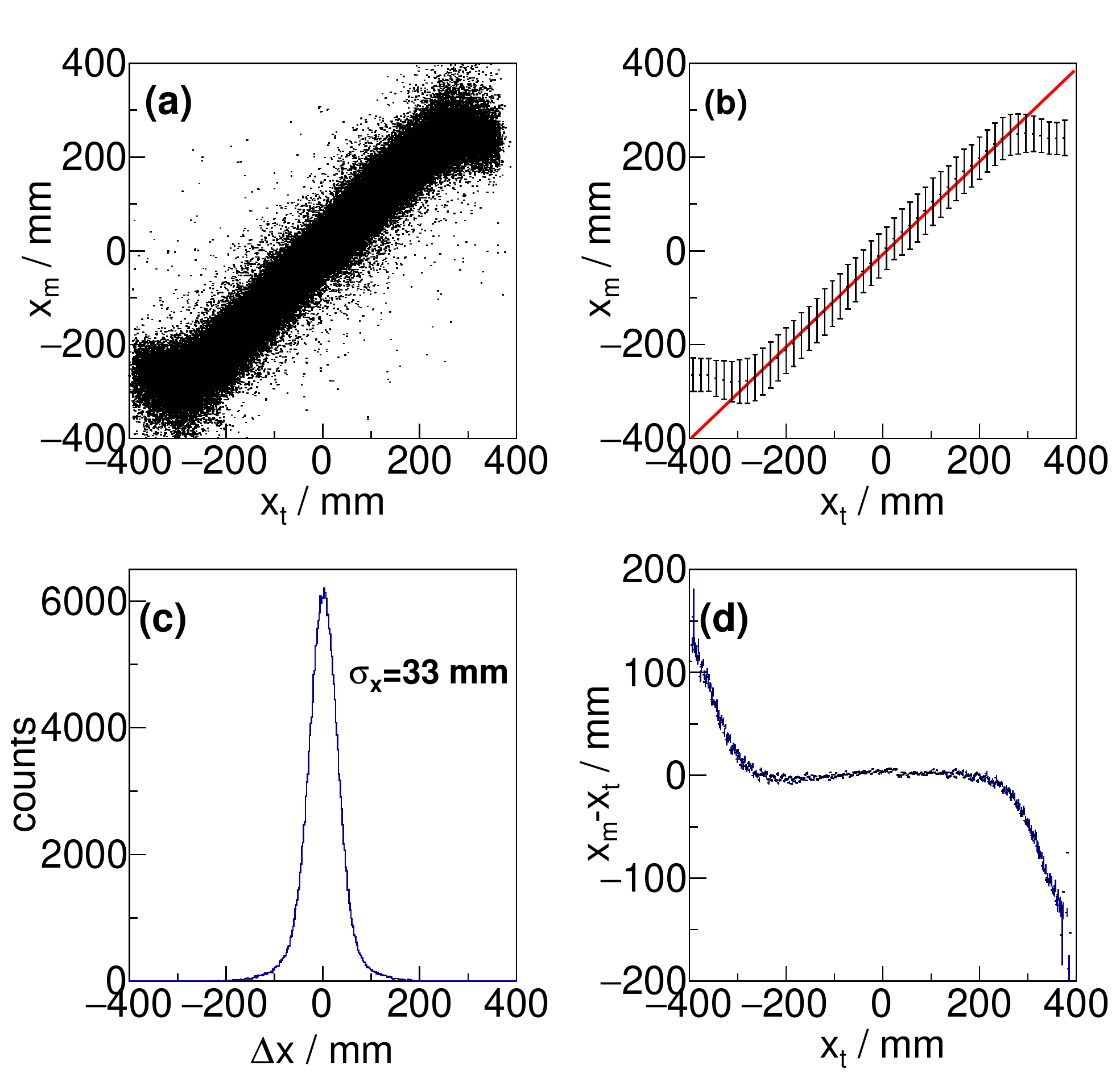}
\figcaption{\label{fig7} The comparison of measured position and real position in x direction. (a) 2-D scatterplot. (b) Position linearity analysis. (c) Position resolution. (d) The deviation between measured and real positions varies with the hit position.}
\end{center}

Fig.~\ref{fig7}a is the value of the measured position vs. the value of the true position along x direction. It shows that the linearity is very good in the center part of the detector but distorts nearby the edges. With the relation of $x_m$ and $x_t$, position non-linearity in x direction can be investigated in the following way. The 2-D spectrum was made N slices (N=50 e.g.) at $x_t$ axis, the center of each slice became a new true position x, labeled as $x_t^\prime$. Events in each slice were projected to $x_m$ axis to get a new measured position x, labeled as $x_m^\prime$. The root
mean square (RMS) detector non-linearity (\%) for x direction was determined as Eq.~\ref{eq5}:
\begin{equation}
\label{eq5}
\delta_x =\frac{\sqrt{\langle(x_m^\prime-x_t^\prime)^2\rangle}}{L}
\end{equation}
where L is the range in x direction. Fig.~\ref{fig7}b shows the position linearity in x direction, the non-linearity $\delta_x$ is 2.8\% for the center part of the detector $x_t$ from -300~mm to 300~mm, and 7.2\% for the whole detector range. As for y direction, the non-linearity $\delta_y$ is 2.9\% for $y_t$ from -300~mm to 300~mm, and 7.8\% for the whole detector range.

The deviation between $x_m$ and $x_t$ is plotted in Fig.~\ref{fig7}c, which leads to a position resolution of $\sigma_x$=33~mm in x direction, and the position resolution in y direction is obtained as $\sigma_y$=35~mm similarly. Therefore, position resolution for the whole detector $\sigma=\sqrt{{\sigma_x}^2+{\sigma_y}^2}$ equals 48~mm.

The deviation varying with the hit position is plotted in Fig.~\ref{fig7}d. The positions reconstructed in the center part of the detector are quite exact, but the deviation goes larger when cosmic rays hit close to the edges, which is also clearly shown in Fig.~\ref{fig8}, a typical position reconstruction for cosmic rays. The positions nearby every edge of the detector are all mis-reconstructed faraway from the edge. It is mainly because that the photocathodes of the PMTs have limited acceptance angle for the scintillation light. When a cosmic ray hit in one position close to some edge, scintillation lights propagated to the neighbouring two PMTs in the shortest path are out of their acceptance angles, only scintillation lights propagated in longer paths can be accepted, which means longer time for these PMTs, but scintillation lights propagated to the other distant PMTs are not affected. As a result, the position reconstruction presents a tendency to bend faraway from the edge. The regions close to the edges are not suitable for position detection. The effective area of the scintillation detector is about 600 $\times$ 600 mm$^2$ in the center as a position sensitive detector.
\begin{center}
\includegraphics[width=6cm]{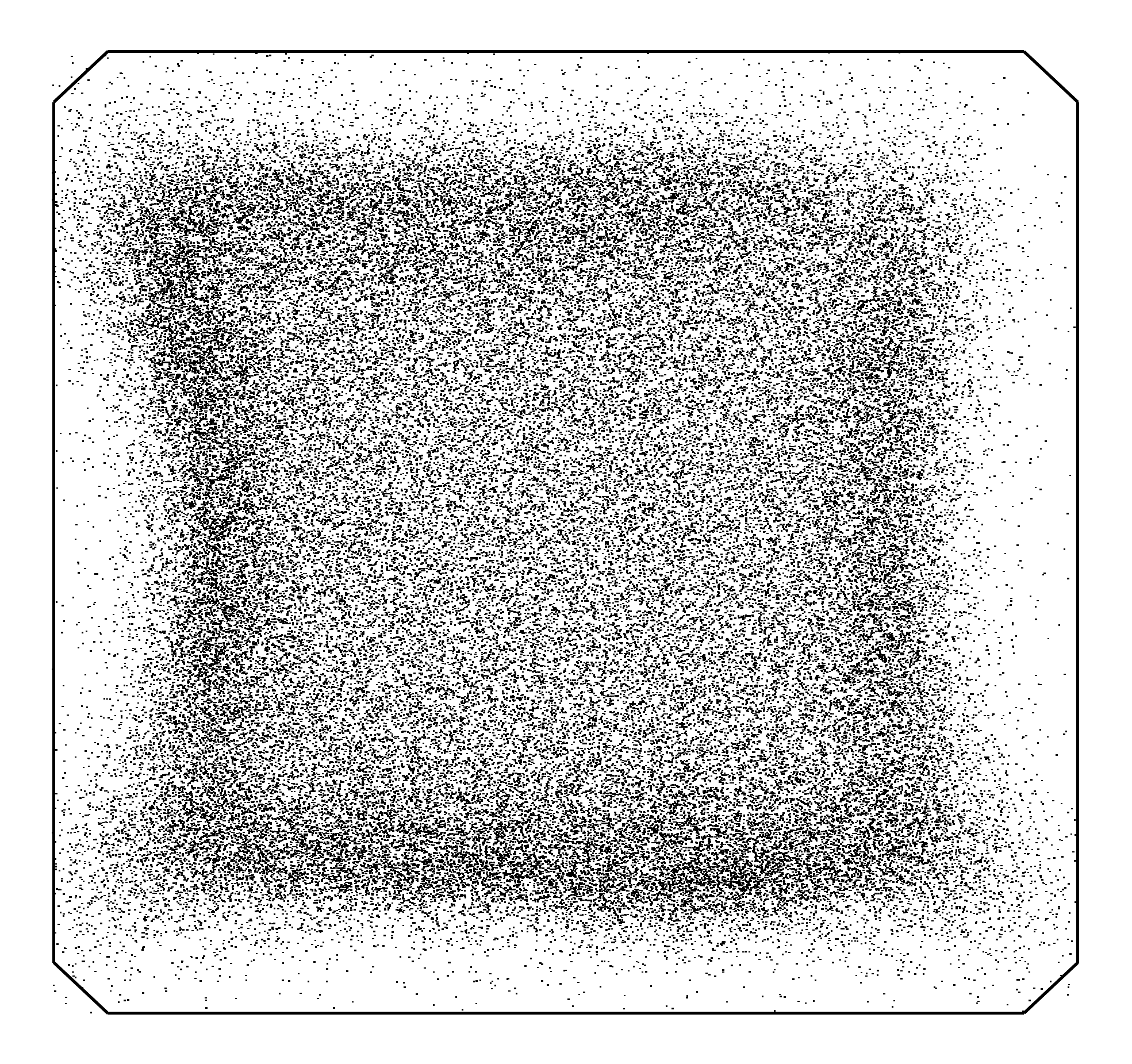}
\figcaption{\label{fig8} Position reconstruction from four corner time for cosmic rays.}
\end{center}

\subsection{Detection efficiency}
The detection efficiency in different positions of the detector was also investigated from the cosmic rays test in Fig.~\ref{fig6}. The whole detector was divided into enough cells according to the true hit positions of the cosmic ray muons, and in each cell the detection efficiency was calculated as $\eta=N/N_t$. Where $N$ means the valid number of cosmic ray muons counted with the tested scintillation detector, and $N_t$ means the total number of cosmic ray muons counted with the upper and lower trigger detectors. One cosmic ray event was considered valid only when four PMT signals of the tested detector were all detected. Fig.~\ref{fig9} shows the detection efficiency of the detector. The detection efficiency is nearly 100\% in the main detector range, but decreases dramatically at the edges.
\begin{center}
\includegraphics[width=6cm]{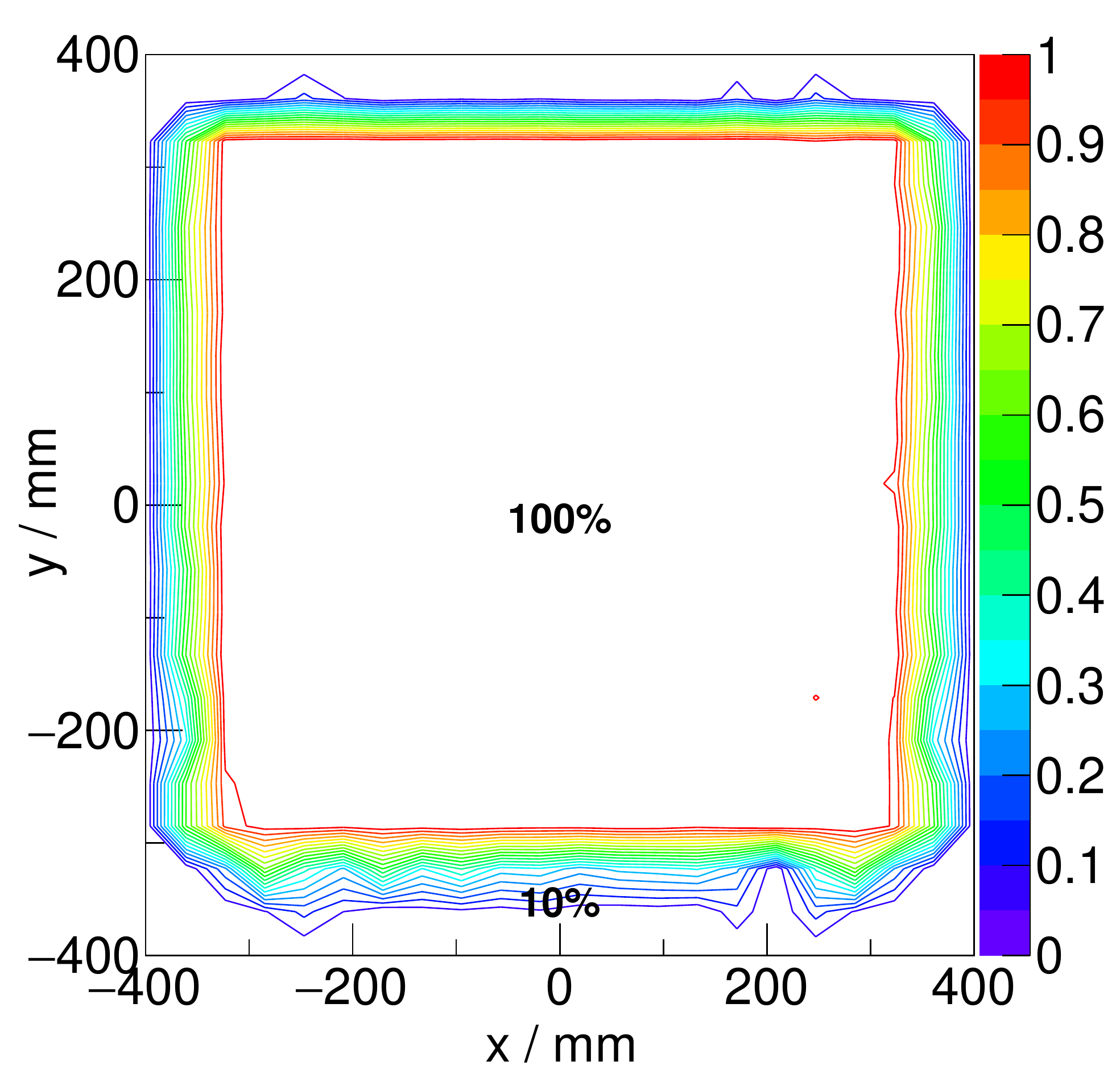}
\figcaption{\label{fig9} Detection efficiency for cosmic rays.}
\end{center}

\section{Summary and discussion}
The construction of a scintillation detector in the size of 760 $\times$ 760 $\times$ 30 mm$^3$ has been described. Time and position performance of the detector have been tested by using $\gamma$ rays and cosmic rays. The time information for one single PMT is position-dependent, but a position-independent effective time $T_{eff}$ has been found, which means the detector can be used as a TOF detector, and the TOF resolution is 236~ps tested with cosmic rays. The position hit by one particle can be reconstructed with the time information from four PMTs in the corners, and the position resolution is 48~mm. The position linearity analysis shows that the non-linearity is about 3\% in the center 600~mm region and more than 7\% for the whole range. The position reconstruction nearby the edges is of large deviation mainly caused by the limited acceptance angle of the PMTs. As a position sensitive detector, the effective area of the scintillation detector is about 600 $\times$ 600~mm$^2$ in the center. The detection efficiency of the detector is close to 100\% in the main detector range, but decreases dramatically at the edges.

Compared to the routine TOF detector made by an array of long scintillation bars, time resolution and position resolution of this detector is slightly worse, but the amount of the readout signals is much less. This detector is much more cost-effective and easier for operation, and could be used in certain nuclear physics experiments.

\acknowledgments{We thank the staff of the gas detector lab for the help on MWDC. .}

\end{multicols}

\vspace{10mm}

\vspace{-1mm}
\centerline{\rule{80mm}{0.1pt}}
\vspace{2mm}

\begin{multicols}{2}

\end{multicols}

\clearpage

\end{document}